\documentclass[12pt]{iopart}
\usepackage{iopams}  

\expandafter\let\csname equation*\endcsname\relax

\expandafter\let\csname endequation*\endcsname\relax

\usepackage{graphicx}%
\usepackage{multirow}%
\usepackage{amssymb,amsfonts}%
\usepackage{amsthm}%
\usepackage{mathrsfs}%
\usepackage[title]{appendix}%
\usepackage{xcolor}%
\usepackage{textcomp}%
\usepackage{manyfoot}%
\usepackage{booktabs}%
\usepackage{algorithm}%
\usepackage{algorithmicx}%
\usepackage{algpseudocode}%
\usepackage{listings}%

\renewcommand*{\bmod}{\mathbin{\%}}

\usepackage{mathtools}

\usepackage{caption}
\usepackage{comment}
\usepackage{enumerate}
\usepackage{amstext,amsopn} 
\usepackage{subfigure}
\usepackage{tcolorbox}

\usepackage{booktabs} 

\usepackage{epigraph}
\usepackage{nicefrac}

\usepackage{subfigure}

\usepackage{extarrows}
\usepackage{url}

\newtheorem{definition}{Definition}

\newtheorem{proposition}{Proposition}

\newtheorem{conjecture}{Conjecture}
\newtheorem{lemma}{Lemma}

\renewcommand{\vec}[1]{\mathbf{#1}}

\newcommand\dmx{\widetilde{x}} 
\newcommand\dmy{\widetilde{y}}  

\newcommand\nx{\chi} 
\newcommand\ny{\phi}  
\newcommand\nv{\omega}  

\newtheorem{theorem}{Theorem}

\begin{document}

\title[Tit-for-tat dynamics and market volatility]{Tit-for-tat dynamics and market volatility}

\author{Simina Br\^anzei}

\address{Purdue University, West Lafayette, IN, US}
\ead{simina.branzei@gmail.com}
\vspace{10pt}
\begin{indented}
\item[]\today 
\end{indented}

\begin{abstract}
	We consider tit-for-tat dynamics in production markets, where there is a set of $n$ players connected via a weighted graph. Each player $i$ can produce an eponymous good using its linear production function, given as input various amounts of goods in the system. 
	In the tit-for-tat dynamic, each player $i$ shares its good with its neighbors in fractions proportional to how much they  helped player $i$'s production in the last round. 
	
	Our contribution is to  characterize the asymptotic behavior of the dynamic as a function of the graph structure, finding that the fortune of a player grows in the long term if and only if the player has a good self loop (i.e. the player works well alone) or works well with at least one other player. 
	We also consider a generalized damped update, where the players may update their strategies with different speeds, and obtain a lower bound on their rate of growth by identifying a function that   gives insight into the behavior of the dynamical system.
	
	The model can capture circular economies, where players use each other's products, and  organizational partnerships, where   fostering long-term growth of an organization hinges on creating relationships in which reciprocal exchanges between the agents in the organization are paramount. 
\end{abstract}

\section{Introduction}
The tit-for-tat strategy has been studied extensively in repeated games, where two agents playing tit-for-tat adjust their behavior to match that of their opponent in the past. For example, in games such as  the repeated Prisoner's dilemma, the choice of a player is between cooperation and retaliation (see, e.g. \cite{FT91}). An agent playing tit-for-tat will cooperate as long as  their opponent cooperates as well, and will subsequently copy the previous action of the other agent. 

More generally, a tit-for-tat strategy prescribes proportional retaliation to avoid escalation of conflict. Axelrod~\cite{axelrod_book} studied tit-for-tat to explain how high levels of cooperation can be achieved in groups of animals and human societies even when each individual player is selfish. Tit-for-tat can also be used to study economic policies such as setting tariffs between countries, where an increase in tariffs by one country is followed by corresponding increases in tariffs by other countries.


In this paper we consider tit-for-tat dynamics in  production markets, where there is a set of $n$ players, each starting with some amount of an eponymous good and a production recipe. The players can produce their own good given as input a bundle of various amounts of goods in the market. 

Our market model is a simple variant of the pioneering model of an expanding economy due to von Neumann \cite{vNeumann46}, where the goods are substitutes (the initial model was for perfect complements), represented through linear production functions \cite{Gale89}.  For linear production, there is a matrix $\vec{v}$ so that player $i$ can make $v_{i,j}$ units of its good given as input one unit of good $j$. The model allows for circular dependencies, which are useful for capturing situations in which people would use each other's products. 

Examples of goods that can be modelled with linear valuations are tea and coffee,  since these resources are substitutes and so can replace each other in consumption.
Linear production can be seen as a first step towards understanding more complex production functions, such as concave ones, which could be approximated by  piecewise linear functions with some finite number of segments. The next figure shows a visual depiction of tit-for-tat dynamics in production markets with linear production, for several randomly generated instances.

\begin{figure}[h!]
	\centering
	\includegraphics[scale=0.49]{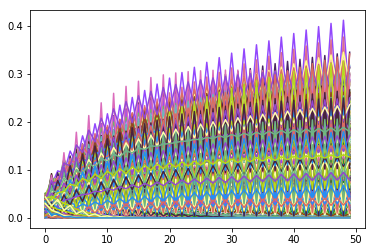}
	\includegraphics[scale=0.49]{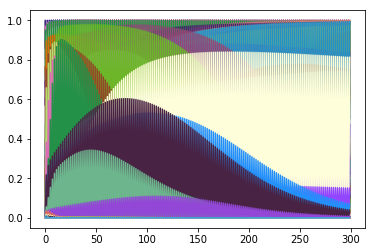}
	\caption{{\textit{Tit-for-tat in two markets with $n=50$ players. Each player $i$ starts with an initial amount $x_i(0)$ of good $i$ and a way of distributing its good given by a vector $\vec{y_i}(0)$, where $y_{i,j}(0)$ is the fraction player $i$ gives to player $j$ (from good $i$) at round $0$. The players repeatedly exchange goods, produce their good from the bundles acquired, and then update the fractions according to the tit-for-tat rule. The picture shows the fractions $y_{j,i}(t)$, which have large oscillations.}}}
\end{figure}

The work of von Neumann as well as most of the literature following it focused on analyzing 
equilibrium-like states where the economy expands by the same  factor as time progresses. 
In contrast, we will focus on dynamics where the players act in a decentralized way, based on the limited information  they  have at the current  time. 

The study of such dynamics is part of a general direction of moving from general equilibrium theory to a computational theory of market dynamics, where the players learn how to play based on partial information that they possess. 
A theoretical understanding of market dynamics could give insight into questions such as: given a model of player learning, will the market grow, is inequality likely to appear? This could later give insight into how to design economic simulators~\cite{salesforce} and explain why, how, or when certain phenomena may appear in real markets. 

Tit-for-tat behavior has been well documented in human and animal interactions~\cite{Axelrod81theevolution}. Tit-for-tat dynamics have been  studied before in exchange markets \cite{WZ07} with linear utilities where the goods have common value (i.e. where each good has the same value to every agent). Our definition of  tit-for-tat dynamics follows the one in \cite{WZ07}, except we analyze it in  production markets rather than exchange markets.

Finally, related   dynamics such as proportional response  and tatonement have also been studied  in various markets, where the sellers bring the same amount of their good each day to the market, but may change the price and the buyers respond by changing what they purchase.

\paragraph{Roadmap to the paper.} In Section \ref{sec:model}  we define the model and state our main results. Section \ref{sec:related_work} contains related work. Section  \ref{sec:tit_for_tat_main} analyzes the tit-for-tat dynamics, while Section \ref{sec:damped_tit_for_tat} analyzes a generalization where the updates are damped.  Section \ref{sec:discussion} discusses  future work.  Section \ref{app:damped} contains the proofs of several lemmas.

\section{Model and Results} \label{sec:model} Let $[n] = \{1, \ldots, n\}$ be a set of players. Each player $i$ can make an eponymous good using its linear production function described by a vector $\vec{v}_i = (v_{i,1}, \ldots, v_{i,n}) \in \mathbb{R}^n$, such that $v_{i,j} \geq 0$ is the number of units of good $i$ the player can make from one unit of good $j$. The player knows their own production function.

A bundle of goods is a vector $\vec{z} = (z_1, \ldots , z_n) \in \mathbb{R}_{+}^n$, where $z_j$ is the quantity of good $j$ in the bundle. Given a
bundle $\vec{z}$, player $i$ can produce $p_i(\vec{z}) = \sum_{j=1}^n v_{i,j} \cdot z_j$ units of  good $i$.

\paragraph{\textbf{Tit-for-tat dynamic.}} In the dynamic we consider, the players share their good with others in hope of receiving greater returns in the future. Then each player
 $(i)$ observes how much others give to them, and  
 $(ii)$ updates to whom and much to give in the next round, proportionally to how much each  player  contributed to their production in the last round. 
Formally, we have the next dynamical system.


\begin{definition}[Tit-for-tat dynamic] \label{def:tft}
	Each player $i$ starts with an amount  $x_i(0) > 0 $ of good $i$, which it allocates among the other players $j \in [n]$ in initial fractions $y_{i,j}(0)$.
	At each time  $t=0,1,2,\ldots$, the next steps take place:
	\begin{itemize}
		\item	\textbf{\emph{Exchange.}} Each player $j $ gives each player $i$ a fraction $y_{j,i}(t)$ of its  good; thus player $i$ receives  $w_{i,j}(t) = y_{j,i}(t) \cdot x_{j}(t)$ units of good $j$. 
		
		\item 	\textbf{\emph{Production.}} Each player $i$ produces its good  using the ingredients from the bundle acquired:
		$
		x_i(t+1) = \sum_{j=1}^n v_{i,j} \cdot w_{i,j}(t)
		$. 
		\item	\textbf{\emph{Strategy update.}} Each player $i$ updates what fraction of their good  to give to each player $j$, according to the contribution of good $j$ in the last round production of player $i$ (\footnote{If $x_i(t+1) =0$, then $y_{i,j}(t+1) = y_{i,j}(t)$. We will focus on non-degenerate initial conditions, which will ensure   $x_i(t) > 0$ for all $t$.}):
		$$
		y_{i,j}(t+1) = \frac{v_{i,j} \cdot w_{i,j}(t)}{x_i(t+1)}\,.
		$$
	\end{itemize}
\end{definition}

\begin{figure}[h!]
	\centering
	\subfigure[Fractions $y_{j,i}(t)$ for all $i,j$ over $ 70$ rounds.]
	{
		\includegraphics[scale=0.5]{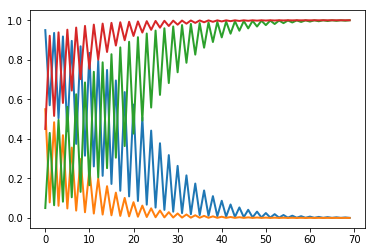}
	}
	\subfigure[Fractions $y_{j,i}(t)$ for all $i,j$ over $250$ rounds.]
	{
		\includegraphics[scale=0.5]{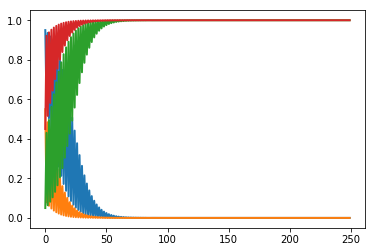}
	}\\
	\subfigure[Amount of good $1$ over time.]
	{
		\includegraphics[scale=0.5]{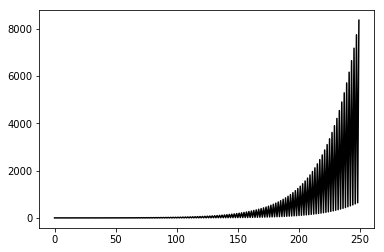}
	}
	\subfigure[Amount of good $2$ over time.]
	{
		\includegraphics[scale=0.5]{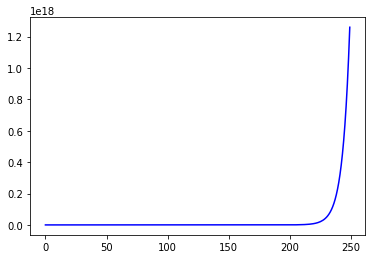}
	}
	\caption{\textit{Tit-for-tat dynamic  in a market with $n=2$ players, where $\vec{v} = [[0.91, 1.186], [0.91, 1.186]]$. Each player $i$ starts with an initial amount $x_i(0)=1$ of good $i$. The initial fractions are $\vec{y}(0)= [[0.95, 0.05], [0.55, 0.45]]$, where $y_{j,i}(t)$ is the fraction received by player $i$ from good $j$ at time $t$. The players repeatedly exchange goods, produce from the bundles acquired, and then update the fractions according to the tit-for-tat rule.}}
	\label{fig:n=2}
\end{figure}

Figure~\ref{fig:n=2}  illustrates the evolution of  fortunes  $x_i(t)$  and investment fractions $y_{i,j}(t)$  in a market with two players.

\paragraph{Initial conditions.} We assume the initial configuration is non-degenerate, which in this context means that for  each player $i \in [n]$ there exists $j \in [n]$ such that $v_{i,j} \cdot v_{j,i} > 0$. Players that do not meet this condition will vanish in a few rounds, meaning the quantity of their good become zero after a constant number of initial rounds.
\medskip  

We also require that the initial amounts are positive, so 
$x_i(0) > 0$ for all $i \in [n]$, and that  players start by sharing a non-zero amount of their good with those players whose goods they value, i.e. $y_{i,j}(0)> 0$ if and only if $v_{i,j} > 0$.

\paragraph{\textbf{Notation.}} We refer to $x_i(t)$ as the \emph{fortune} of player $i$ at time $t$.
The fortune \emph{grows} if $\lim_{T \to \infty} x_i(T) = \infty$ and \emph{vanishes} if $\lim_{T \to \infty} x_i(T) =0$, respectively. 

\begin{theorem} \label{thm:1}
	Let $i \in [n]$ be any player. Then 
	\begin{align}
		x_i(t) \in \Theta\left(\max_{j \in [n]}\left(v_{i,j}  v_{j,i}\right)^{\frac{t}{2}}\right)\,. \notag 
	\end{align}
	The constant in $\Theta$ depends on the matrix $\vec{v}$ and the initial configuration.
	Moreover,
	$\lim_{t \to \infty} y_{i,\ell}(t) = 0$ for each player $\ell$ with $v_{i,\ell}  v_{\ell,i} < \max_{j \in [n]} v_{i,j} v_{j,i}$. 
\end{theorem}
In other words, the fortune of player $i$ grows at a rate given by the geometric mean of the best cycle of length at most two that player $i$ is part of. Moreover, in the limit player $i$ only shares its good with players on such best cycles. 
This also implies that if player $i$ has a bad self loop (i.e.  $v_{i,i} < 1$) \emph{and} all the cycles with its neighbours $j$ are bad (i.e.  $v_{i,j} v_{j,i} < 1$ for all $j \in [n]$), then  the fortune of player $i$ will vanish in the limit \footnote{This property  holds even if the graph with weights $\vec{v}$ may have longer  production cycles on which the product of coefficients is strictly more than $1$. Such a cycle $C$, with $\prod_{(i,j) \in C} v_{i,j} > 1$, could be used by a  central planner to ensure growth of the economy in the long term by having the players coordinate sending their good to the next player in $C$.}

\medskip  

We also  characterize the fortunes and fractions of the players at each point in time.


\paragraph{Damped tit-for-tat.} Next we consider a generalization of tit-for-tat where the updates are damped. We  will call the variables in the damped dynamic $\dmx_i(t)$ and $\dmy_{i,j}(t)$.

The only change from Definition~\ref{def:tft}  is that in the strategy update, each player $i$ has a number $\epsilon_i \in (0,1]$ such that $$\dmy_{i,j}(t+1) = \epsilon_i \cdot \frac{v_{i,j} \dmy_{j,i}(t) \dmx_j(t)}{\dmx_i(t+1)} + (1-\epsilon_i) \dmy_{i,j}(t)\,.$$

For the damped dynamic we show that the fortune of each player $i$ grows at a rate at least as high as the geometric mean of the best cycle of length at most two that player $i$ is part of. The precise statement is for the last two iterates of the amount of player  $i$.

\begin{theorem}[Damped tit-for-tat] \label{thm:2}
	Let $i \in [n]$ be any player. Then 
	$$
	\dmx_i(t) + \dmx_i(t+1) \in \Omega\left(\max_{j \in [n]}\left(v_{i,j} v_{j,i}\right)^{\frac{t}{2}}\right) \forall t \in \mathbb{N}\,.
	$$The constant in $\Omega$ depends on the matrix $\vec{v}$, the initial configuration, and the vector $\epsilon$.
\end{theorem}

\section{Related Work} \label{sec:related_work}

Tit-for-tat dynamics were studied before in pure exchange markets (i.e. without production) in~\cite{WZ07}, where there is a set of players, such that each player $i$ owns one unit of good $i$. The players have linear valuations given by a matrix $\vec{a}$, where $a_{i,j}$ is the value of player $i$ for a unit of good $j$.
Each player repeatedly brings to the market one unit of a good in order to exchange it for other goods that are potentially more valuable.
In tit-for-tat each player decides in what fractions to split its good among its neighbours, and then updates the fractions in the next round proportionally to the utility received from each neighbour. 
The main result in~\cite{WZ07} is that when the valuations are symmetric (see \footnote{Symmetric valuations are such that the value of any player $i$ for a good $j$ is $a_{i,j} = v_j$.}),  the dynamic converges to market equilibria for all non-degenerate starting configurations. 

A related dynamic studied in markets is proportional response, in which every player starts with some amount of good and an initial budget of money that it can use for acquiring goods. In each round, the players split their budget into bids on the goods, then each good is allocated to each player in proportion to the bid amount and the seller collects the money made from selling. 
In the exchange economy, each player updates their bids on the goods in proportion to the contribution of each good in their utility; there, the dynamic converges to market equilibria~\cite{BDR19} for any economy with additive valuations. \cite{CHSV24} studied a related market game over time, where the agents bring goods to the market and the allocations are computed via the proportional sharing mechanism~\cite{feldman2005price}. In each round, the (unique) equilibrium resource allocation becomes the initial endowments for the next round. See \cite{Branzei21} for a survey on proportional response dynamics in exchange economies.

In Fisher markets, proportional response converges to market equilibria, which was shown for additive valuations in~\cite{Zhang11}. \cite{CCT18} show that 
proportional response dynamics converge to market equilibria for the whole range of constant elasticity of substitution (CES) utilities,  including \emph{complements}, with linear 
utilities on one extreme and \emph{Leontief} utilities on the other extreme. \cite{CHN19} show that 
the dynamics stays close to equilibrium even when the market parameters are changing slowly over time, once again 
for CES utilities. \cite{kolumbus2023asynchronous} studied asynchronous proportional response dynamics in Fisher markets.

In Fisher markets with additive valuations, the proportional response dynamic is equivalent to gradient descent~\cite{BDX11}. \cite{CLP21} studied learning dynamics in several production economies, such as blockchain
mining, peer-to-peer file sharing and crowdsourcing. They also study Fisher markets and show that  proportional response dynamics converges to market equilibria when the players have  quasi-linear utilities. \cite{YK20} studied linear, quasi-linear, and Leontief utilities in Fisher markets. They show that mirror descent applied to a new convex program achieves sublinear last-iterate convergence and yields a form of proportional response dynamics.

In production markets, where the players are endowed with linear production functions, the proportional response dynamic was shown in~\cite{BMN18} to
lead to universal growth of the market, where the amount of goods produced  
grows over time (whenever growth is possible), but also to growing inequality between the players on the most efficient 
production cycle and the rest. In particular, the dynamic learns through local interactions a global feature of the 
graph with weights $\vec{v}$, namely the cycle with the highest geometric mean.

We note  a difference between the proportional response dynamic from ~\cite{BMN18} and the tit-for-tat dynamic we focus on. In the proportional response dynamics, if the graph of valuations has a cycle with geometric mean greater than 1 (see \footnote{That is, if there is a cycle $C$ in the graph with nodes $[n]$ and weights $v_{i,j}$ such that $\prod_{(i,j) \in C} v_{i,j} > 1$.}), then the fortunes of all the players on that cycle grow in the long term. A crucial difference is that the  proportional response dynamics requires the players to have money;  players submit bids on goods that they like, the sellers of the goods collect the money from selling, and use the resulting budget in their own bidding in the next period. 
On the other hand, the tit-for-tat dynamic from our setting here works without money, yet it still leads to some degree of cooperation between the players, though on a smaller scale.

Another central dynamic studied in markets is tatonement, which has been studied in a series of papers on markets for players with additive, CES (constant elasticity of substitution), and Leontief valuations~\cite{cole2008fast,CCT18,cheung2019tatonnement,GoktasZG23}. Tatonement does not prescribe how allocations may be formed, but defines how prices are adapted depending on the demand in the previous round.
The Shapley-Shubik game~\cite{shapley1977trade}, which forms the basis of the proportional response dynamic, was studied in static Fisher and exchange settings for its equilibrium qualities~\cite{johari2004efficiency,feldman2005price}. 

 Markov exchange economies were studied in \cite{markov_exchange_reinforcement} from the point of view of multi-agent reinforcement learning. There, there is a set of agents that  respond myopically at each step while aiming to maximize their own utility, and a central planner whose goal is to steer the system towards states that maximize the social welfare.

The model we consider is also reminiscent of replicator dynamics~\cite{taylor_replicator,SK83} and the Lotka-Volterra model from biology~\cite{Lotka10,volterra28,W78}, where the matrix $\vec{v}$ captures interactions between  different animal and plant species, which can help each other survive or destroy each other (e.g. wolves reduce the population of deer, while edible plants help the deer population grow). The population of each species can grow or vanish, depending on the interactions with others.

\section{Tit-for-Tat Dynamics} \label{sec:tit_for_tat_main}

We start by analyzing the dynamic from Definition~\ref{def:tft}.  This will help build the intuition that exchange  essentially takes place along cycles of length at most $2$, which will then  form the basis of the analysis for the damped system.

\begin{lemma}\label{lem:greedy_product}
	For all $i,j \in [n]$ and all $t \in \mathbb{N}_{\geq 1}$,  we have 
	\begin{align} \label{eq:basic_identity_amount_prod_fraction}
		x_i(t)  y_{i,j}(t) = (v_{i,j}  v_{j,i})^{\lfloor \frac{t}{2} \rfloor}  x_i(r)  y_{i,j}(r), \; \; \; \text{where $r = t \bmod{2}$.}
	\end{align}
\end{lemma}
\begin{proof}
	We consider a few cases.  
	\paragraph{\textbf{Case 1:}} $v_{i,j} \cdot v_{j,i} > 0$. The update in Definition \ref{def:tft} gives 
	\begin{align}  \label{eq:simple_identities_y_ij}
		\begin{cases}
			\frac{y_{i,j}(t)}{y_{j,i}(t-1)}  =  \frac{v_{i,j} \cdot x_j(t-1)}{x_i(t)} \; \; \forall  t \geq 1  \\[5pt]
			\frac{y_{j,i}(t-1)}{y_{i,j}(t-2)}  =  \frac{v_{j,i} \cdot x_i(t-2)}{x_j(t-1)} \; \; \forall  t \geq 2\,.
		\end{cases}
	\end{align} 
	
	Multiplying the  identities in \eqref{eq:simple_identities_y_ij} implies  
	\begin{align} \label{eq:useful_id}
		& \frac{y_{i,j}(t)}{y_{i,j}(t-2)} = (v_{i,j}  v_{j,i}) \cdot \frac{x_i(t-2)}{x_i(t)}\;\; \forall t \geq 2 \,.
	\end{align}
	Let $r = t \bmod{2}$. 
	Taking the product of  (\ref{eq:useful_id}) for a range of even and odd time steps, respectively, we obtain: 
	\begin{align} \label{eq:odd_id}
		& \; \; \; \prod_{\ell=1}^{\lfloor {t}/{2} \rfloor } \frac{y_{i,j}(2\ell+r)}{y_{i,j}(2\ell-2+r)} = \prod_{\ell=1}^{\lfloor {t}/{2} \rfloor } v_{i,j}  v_{j,i} \cdot  \frac{x_{i}(2\ell-2+r)}{x_{i}(2\ell+r)}    \notag \\
		& \implies  {y_{i,j}(t)}/{y_{i,j}(r)} = (v_{i,j}  v_{j,i})^{\lfloor {t}/{2} \rfloor }  {x_i(r)}/{x_i(t)}.
	\end{align}
	Identity  (\ref{eq:odd_id}) yields   \eqref{eq:basic_identity_amount_prod_fraction}, as required by the lemma.
	\paragraph{\textbf{Case 2:}} $v_{i,j} > 0$ and $v_{j,i} =0$. Then $y_{i,j}(0) > 0$ and $y_{j,i}(0) = 0 $ by definition of the initial configuration as non-degenerate. The update rule rules gives 
	\begin{align} 
		\begin{cases} 
			y_{i,j}(1) = {v_{i,j} y_{j,i}(0)  x_j(0)}/{x_i(1)} = 0; \\[2pt]
			y_{j,i}(1) = {v_{j,i} y_{i,j}(0) x_i(0)}/{x_j(1)} = 0\,. \notag 
		\end{cases} 
	\end{align} 
	
	Then  $y_{i,j}(t) = y_{j,i}(t) = 0$ $\forall t \in \mathbb{N}_{\geq 1}$, and so equation (\ref{eq:basic_identity_amount_prod_fraction}) holds.
	The case  $v_{j,i} > 0$ and $v_{i,j} = 0$ is symmetric.
	
	\medskip 
	\paragraph{\textbf{Case 3:}} $v_{i,j} = v_{j,i} = 0$. Since the initial configuration is non-degenerate, we have   $y_{i,j}(0) = y_{j,i}(0) = 0$ . Then  $y_{i,j}(t) = y_{j,i}(t) = 0$ $\forall t \in \mathbb{N}$, so equation (\ref{eq:basic_identity_amount_prod_fraction})  holds. 
\end{proof}

\begin{proposition} \label{thm:char}
	The fortune of each player $i \in [n]$  at time $t \in \mathbb{N}_{\geq 1}$ is:
	\begin{align}
	x_i(t) = 
	\begin{cases}
		\begin{matrix}
			\sum\limits_{j=1}^{n} \left( v_{i,j}  v_{j,i}\right)^{ \frac{t}{2}}  x_i(0)  y_{i,j}(0) \; \; \; \text{ if } t \text{ even}; 
		\end{matrix}
		\\[15pt]
		\begin{matrix}
			\sum\limits_{j=1}^n  \left( v_{i,j} \right)^{\lceil \frac{t}{2} \rceil}  \left( v_{j,i} \right)^{\lfloor \frac{t}{2} \rfloor }   x_j(0)  y_{j,i}(0) \; \; \; \text{ if } t \text{ odd} \,.  
		\end{matrix}
	\end{cases} \notag 
	\end{align}
\end{proposition}
\begin{proof}
	Let $r = t \bmod{2}$.	Recalling that  $\sum_{j=1}^n y_{i,j}(t) = 1$, take the sum of 
    (\ref{eq:basic_identity_amount_prod_fraction})  for all $j$:
	\begin{align} \label{eq:odd_ub}
		& \sum_{j=1}^n x_{i}(t)  y_{i,j}(t) = \sum_{j=1}^n  (v_{i,j}  v_{j,i})^{\lfloor \frac{t}{2} \rfloor }  x_i(r)  y_{i,j}(r)     \notag \\ 
		& \iff  x_{i}(t) = \sum_{j=1}^n  (v_{i,j}  v_{j,i})^{\lfloor \frac{t}{2} \rfloor } x_i(r)  y_{i,j}(r)   \,. 
	\end{align}
	For  $t$ even, equation \eqref{eq:odd_ub} represents the   required identity.
	
	For $t$ odd, recall   $ x_i(1)  y_{i,j}(1)   = v_{i,j}  x_j(0)  y_{j,i}(0) $ ($\dagger$) by definition of the update rule for $y_{i,j}(1)$.  Substituting ($\dagger$) in (\ref{eq:odd_ub}) gives:
	\begin{align} 
		x_{i}(t) & = \sum_{j=1}^n  (v_{i,j}  v_{j,i})^{\lfloor \frac{t}{2} \rfloor }   x_i(1)   y_{i,j}(1)   = \sum_{j=1}^n  (v_{i,j}  v_{j,i})^{\lfloor \frac{t}{2} \rfloor }  v_{i,j}  \cdot x_j(0)   y_{j,i}(0) \,. \notag 
	\end{align}
	This completes the characterization of the fortunes.
	%
\end{proof}

Next we   characterize the investment fractions $y_{i,j}(t)$.



\begin{proposition} \label{cor:fractions_characterization}
	For each  $i,j \in [n]$ and round $t \in \mathbb{N}_{\geq 1}$, the fraction $y_{i,j}(t)$ is:
	\begin{align}
		y_{i,j}(t) =  
		\begin{cases} 
			\frac{\left( v_{i,j}  v_{j,i} \right)^{{t}/{2}}   x_i(0)  y_{i,j}(0) }{\sum\limits_{k=1}^{n}    \left( v_{i,k}  v_{k,i} \right)^{{t}/{2}} x_i(0)  y_{i,k}(0) }  
			&
			\text{ if } t \text{ even;}
			\\[20pt]
			\frac{  \left( v_{i,j} \right)^{\lceil {t}/{2} \rceil}  \left( v_{j,i} \right)^{\lfloor {t}/{2} \rfloor } x_j(0)  y_{j,i}(0)  }{\sum\limits_{k=1}^n   \left( v_{i,k} \right)^{\lceil {t}/{2} \rceil} \left(  v_{k,i} \right)^{\lfloor {t}/{2} \rfloor } x_k(0)  y_{k,i}(0)}
			&
			\text{ if } t \text{ odd.}
		\end{cases} \notag 
	\end{align}
\end{proposition}
\begin{proof}
	Let $r = t \bmod{2}$. 
	From Lemma \ref{lem:greedy_product} we obtain 
	\begin{align} \label{eq:simple_formula_y_ij}
		y_{i,j}(t) = (v_{i,j}  v_{j,i})^{\lfloor \frac{t}{2} \rfloor}  x_i(r)  y_{i,j}(r)/x_i(t)\,.
	\end{align} Substituting the formula for $x_i(t)$ from Proposition~\ref{thm:char} in   (\ref{eq:simple_formula_y_ij}) gives the   required identity for $y_{i,j}(t)$ when $t$ is even. 
 
	For odd $t$,   the update rule for $y_{i,j}(1)$ yields  
	\begin{align} x_i(1) y_{i,j}(1)  = v_{i,j}  x_j(0) y_{j,i}(0) \;  (\ddagger)\,.  \notag 
 \end{align}
	Substituting ($\ddagger$) in equation (\ref{eq:simple_formula_y_ij}) gives: $y_{i,j}(t) = {(v_{i,j}  v_{j,i})^{\lfloor {t}/{2} \rfloor}  x_i(1)  y_{i,j}(1)}/{x_i(t)} .$ Using  Proposition~\ref{thm:char}  implies the required formula for $y_{i,j}(t)$ when $t$ odd as well.
\end{proof}

\begin{lemma} \label{lem:fractions_limit_behavior}
	For each player $i$, 
	let $S_i = \{j \in [n] \mid v_{i,j}  v_{j,i} = \max_{k \in [n] }v_{i,k}  v_{k,i}\}\,.$
	Then for all $k \not \in S_i$, we have $\lim_{t \to \infty} y_{i,k}(t) = 0$. 
	For all $k \in S_i$, we have  
	\begin{align}
		\lim\limits_{t \to \infty} y_{i,k}(2t) & =
		\frac{ y_{i,k}(0)}{\sum\limits_{\ell \in S_i}   y_{i,\ell}(0)}    \; \; \; \;   \text{and}   \; \; \; \;  
		\lim\limits_{t \to \infty} y_{i,k}(2t+1) =
		\frac{v_{i,k} \cdot  y_{k,i}(0)  x_k(0)}{\sum\limits_{\ell \in S_i}v_{i,\ell}  \cdot  y_{\ell,i}(0)  x_{\ell}(0)}\,.
	\end{align}
\end{lemma}
\begin{proof}Let $k \in [n]$. Define $r = t \bmod{2}$. We consider two cases, where $k \in S_i$ and $k \not \in S_i$ 
	
	\paragraph{\textbf{Case $k \in S_i$.}} Equation (\ref{eq:simple_formula_y_ij}) from Proposition~\ref{cor:fractions_characterization} gives 
	\begin{align} \label{eq:almost_limit_formula_yij}
		y_{i,k}(t) & = \frac{\left(v_{i,k} v_{k,i}\right)^{\lfloor \frac{t}{2} \rfloor } x_i(r) y_{i,k}(r)}{\sum\limits_{\ell =1}^{n} \left(v_{i,\ell} v_{\ell,i}\right)^{\lfloor \frac{t}{2} \rfloor } x_i(r) y_{i,\ell}(r)} \,.
	\end{align}
	
	For all $k  \in S_i$ and  $\ell \not \in S_i$ we have  $\frac{v_{i,\ell}v_{\ell,i}}{v_{i,k}v_{k,i} } < 1$,  so $$\left(\frac{v_{i,\ell}v_{\ell,i}}{v_{i,k}v_{k,i} }\right)^{t} \xrightarrow[t \to \infty]{}  0 \; \; \; (\mathsection)\,.$$
	
For $r=0$, using ($\mathsection$) in  equation \eqref{eq:almost_limit_formula_yij}  gives: 
	$$\lim_{t \to \infty} y_{i,k}(2t) = \frac{x_i(0)y_{i,k}(0)}{\sum\limits_{\ell \in S_i} x_i(0) y_{i,\ell}(0)} = \frac{y_{i,k}(0)}{\sum\limits_{\ell \in S_i} y_{i,\ell}(0)}\,.$$
	For $r=1$,  the identity $x_i(1) y_{i,k}(1) = v_{i,k} \cdot y_{k,i}(0)  x_k(0)$ combined with ($\mathsection$) gives  
	\begin{align}
		\lim_{t \to \infty} y_{i,k}(2t+1) &= \frac{x_i(1)y_{i,k}(1)}{\sum\limits_{\ell \in S_i} x_i(1) y_{i,\ell}(1)}   = \frac{v_{i,k} \cdot y_{k,i}(0)  x_k(0)}{\sum\limits_{\ell \in S_i}v_{i,\ell} \cdot y_{\ell,i}(0)  x_{\ell}(0)} \,. \notag 
	\end{align}
	\paragraph{\textbf{Case  $k \not \in S_i$.}} Summing   equation (\ref{eq:almost_limit_formula_yij}) gives $\lim_{t \to \infty} \sum_{k \in S_i} y_{i,k}(t) = 1$. Then  $\lim_{t \to \infty} y_{i,k}(t) =0$ for all $k \not \in S_i$ as required.
\end{proof}

\begin{lemma} \label{lem:amounts_asymptotic_ratio}
	For each player $i$, there exist constants $c_i, d_i > 0$, which depend on the matrix $\vec{v}$ and the initial configuration, such that:
	$$
	c_i  \cdot \max_{j \in [n]}\bigl(v_{i,j} v_{j,i}\bigr)^{\frac{t}{2}} \leq x_i(t) \leq d_i  \cdot \max_{j \in [n]} \bigl(v_{i,j} v_{j,i}\bigr)^{\frac{t}{2}} \; \; \; \forall t \in \mathbb{N}\,.
	$$
\end{lemma}
\begin{proof} Let $ r = t \bmod{2}.$
	Equation (\ref{eq:odd_ub})  gives 
	$ x_{i}(t) = \sum_{k=1}^n  \bigl(v_{i,k}  v_{k,i}\bigr)^{\lfloor \frac{t}{2} \rfloor } x_i(r)  y_{i,k}(r)$ ($\diamond$). We can upper bound the fortune of player $i$ at  time $t$ as follows:		\begin{align}  
		x_{i}(t) & = \sum_{k=1}^n  (v_{i,k}  v_{k,i})^{\lfloor \frac{t}{2} \rfloor } x_i(r)  y_{i,k}(r) \notag \\
		& \leq \sum_{k=1}^{n} \bigl(\max_{j\in [n]} v_{i,j}  v_{j,i} \bigr)^{\lfloor \frac{t}{2} \rfloor } x_i(r)  y_{i,k}(r)   =  x_i(r) \bigl(\max_{j\in [n]} v_{i,j}  v_{j,i} \bigr)^{\lfloor \frac{t}{2} \rfloor }  \notag \\
		& \leq \max_{r \in \{0,1\}} x_i(r) \left( 1 + \frac{1}{\max\limits_{j \in [n]} v_{i,j} v_{j,i}}\right) \bigl(\max_{j\in [n]} v_{i,j}  v_{j,i} \bigr)^{\frac{t}{2}}\,. \notag 
	\end{align}
	Setting $d_i = \max_{r \in \{0,1\}} x_i(r)   \Bigl( 1 + \frac{1}{\max_{j \in [n]} v_{i,j} v_{j,i}}\Bigr)$ gives the required upper bound. 
	
	\medskip 
	
	For the lower bound, using ($\diamond$) gives:
	\begin{align}
		x_{i}(t) & = \sum_{k=1}^n  \bigl(v_{i,k}  v_{k,i}\bigr)^{\lfloor \frac{t}{2} \rfloor } x_i(r)  y_{i,k}(r) \notag \\
		& \geq \max_{j\in [n]} \bigl( v_{i,j}  v_{j,i} \bigr)^{\lfloor \frac{t}{2} \rfloor} \cdot \min_{r \in \{0,1\}; k \in [n]}  x_i(r)   y_{i,k}(r) \notag \\
		& \geq c_i \cdot \max_{j \in [n]}  \bigl(v_{i,j} v_{j,i}\bigr)^{\frac{t}{2}}, \qquad \; \; \mbox{ where } c_i = \frac{\min\limits_{r \in \{0,1\}; k \in [n]}  x_i(r)   y_{i,k}(r) }{1 + \max\limits_{k \in [n]} \sqrt{v_{i,k} v_{k,i}}}\,.  
	\end{align} 
	The value $c_i$ depends only on the initial configuration and the values $v_{i,j}$, but is independent of time.
	This completes the proof.
\end{proof}


\medskip 

\begin{proof}[Proof of Theorem~\ref{thm:1}]
	The proof follows by combining    Lemma~\ref{lem:fractions_limit_behavior} and Lemma~\ref{lem:amounts_asymptotic_ratio}. 
\end{proof}

\medskip

\begin{figure}[h!]
	\centering
	\subfigure[Market with $3$ players]
	{
		\includegraphics[scale=0.55]{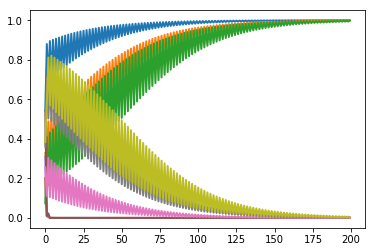}
	}
	\subfigure[Market with $5$ players.]
	{
		\includegraphics[scale=0.55]{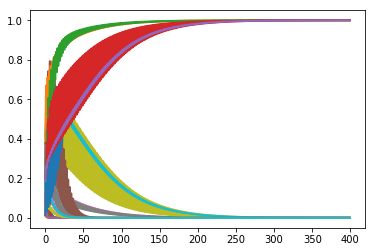}
	}
	%
	\subfigure[Market with $25$ players.]
	{
		\includegraphics[scale=0.55]{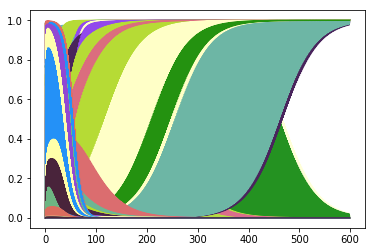}	
	}
	\subfigure[Market with $50$ players.]
	{
		\includegraphics[scale=0.55]{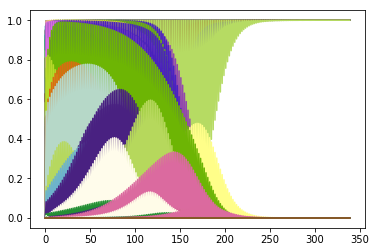}	
	}
	\caption{\textit{Fractions $y_{j,i}(t)$, for all $i,j$, in four random markets with (a) $n=3$ players,  (b) $n=5$ players, (c) $n=25$ players, and (d) $n=50$ players over time, for the  dynamic of Definition~\ref{def:tft}. Each color represents one trajectory of $y_{j,i}(t)$ for some pair $(i,j)$. The fractions have  large fluctuations before converging and so appear as a region.}}
	\label{fig:more_examples_greedy_fractions}
\end{figure}

Figure~\ref{fig:more_examples_greedy_fractions} illustrates the evolution of the fractions $y_{i,j}(t)$ in  several randomly generated markets. 

\section{Damped Tit-for-Tat Dynamics} \label{sec:damped_tit_for_tat}

We consider a more general version of the tit-for-tat update,
where the updates are damped. In the damped dynamic,  each player $i$ updates its strategy by giving weight $\epsilon_i$ to  the result of the last round and weight $1-\epsilon_i$ to its  strategy from before. The dynamic of Definition~\ref{def:tft} is a special case of the damped dynamic with $\epsilon_i = 1$ for all $i \in [n]$.


\begin{definition}[Damped tit-for-tat dynamic] \label{def:tft_savings_lazy}
	Each player $i$ starts with an amount  $\dmx_i(0) > 0 $ of good $i$, which it allocates among the other players $j \in N$ in initial fractions $\dmy_{i,j}(0)$. The player also has a number $\epsilon_i \in (0,1]$.
	At each time  $t=0,1,2,\ldots $, the next steps take place:
	\begin{itemize}
		\item 	\textbf{\emph{Exchange.}} Each player $j $ gives each player $i$ a fraction $\dmy_{j,i}(t)$ of its  good; thus player $i$ gets  $\widetilde{w}_{i,j}(t) = \dmy_{j,i}(t) \cdot \dmx_{j}(t)$ units of  good $j$. 
		
		\item 	\textbf{\emph{Production.}} Each player $i$ produces  its good using the ingredients from the bundle acquired:
		$
		\dmx_i(t+1) = \sum_{j=1}^n v_{i,j} \cdot \widetilde{w}_{i,j}(t)
		$. 
		\item 	\textbf{\emph{Strategy update.}} Each player $i$ updates\footnote{If $\dmx_i(t+1) =0$, then $\dmy_{i,j}(t+1) = \dmy_{i,j}(t)$. We  focus on non-degenerate initial conditions, which will ensure   $\dmx_i(t) > 0$ for all $t$.} how much to give to player $j$:
		$$
		\dmy_{i,j}(t+1) = \epsilon_i \cdot  \left(\frac{v_{i,j} \cdot \widetilde{w}_{i,j}(t)}{\dmx_i(t+1)}\right) + (1-\epsilon_i)\cdot \dmy_{i,j}(t)\,.
		$$
	\end{itemize}
\end{definition}

\begin{figure}[h!]
	\centering
	\includegraphics[scale=0.55]{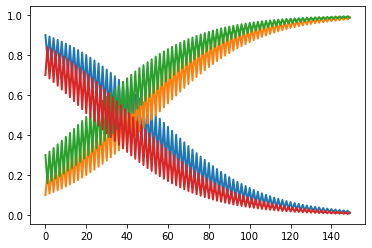}
	\includegraphics[scale=0.55]{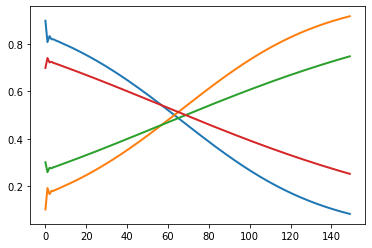}
	\caption{{\textit{Tit-for-tat in a market with two players, where $\vec{v} = [[1.125, 0.915], [1.5, 1.125]]$, the initial fractions are $ [[0.9, 0.1], [0.3, 0.7]]$, and the initial quantities are  $[0.01, 0.01]$. The left figure shows the non-damped dynamic of Definition~\ref{def:tft}, i.e. where $\mathbf{\epsilon} = [1, 1]$. The right figure shows the damped dynamic on the same market starting from the same initial configuration, where $\vec{\epsilon} = [0.8, 0.3]$.}}}
	\label{fig:damped_greedy_comparison}
\end{figure}

Figure~\ref{fig:damped_greedy_comparison} shows an example,  comparing the evolution of the fractions in the damped dynamic with the non-damped one (i.e. where $\epsilon_i=1$ for all $i$).

For the purpose of the analysis, it will be useful to consider a normalized version of the market, where the amounts are rescaled so that the initial amounts sum up to $1$ and the weights $v_{i,j}$ are normalized by dividing them by the largest edge weight. 

More formally, let 
\begin{align} \label{eq:rescaling_damped}
\begin{cases}
	\nx_{i}(0)  = \frac{\dmx_i(0)}{\sigma}, & \forall i \in [n], \mbox{ where } \sigma = \sum_{k=1}^n \dmx_k(0)\,.  \\
	\omega_{i,j} = \frac{v_{i,j}}{\delta}, & \forall i,j \in [n], \mbox{ where } \delta = \max_{k,\ell \in [n]} v_{k,\ell} \,.   \\
	\ny_{i,j}(0)  = \dmy_{i,j}(0) & \forall i,j \in [n] \,.
 \end{cases}
\end{align}


\begin{lemma} \label{lem:normalized_system_relation}
	Suppose 
	\begin{itemize}
		\item $(\vec{\dmx}(t), \vec{\dmy}(t))_{t \geq 0}$ are the variables of the dynamical system of Definition~\ref{def:tft_savings_lazy} with matrix $\vec{v}$, vector $\vec{\epsilon}$, initial amounts $\vec{\dmx}(0)$, and initial fractions $\vec{\dmy}(0)$.
		\item $(\vec{\nx}(t), \ny(t))_{t\geq 0}$ are the variables of the dynamical system of Definition~\ref{def:tft_savings_lazy} with matrix $\vec{\omega}$, vector $\vec{\epsilon}$, initial amounts $\vec{\nx}(0)$, and initial fractions $\vec{\ny}(0)$ as defined in (\ref{eq:rescaling_damped}). 
	\end{itemize} 
	Then $\nx_i(t) = \frac{\dmx_i(t)}{\sigma \delta^{t}}$, $\ny_{i,j}(t) = \dmy_{i,j}(t)$, $\nv_{i,j} \in [0,1]$, and $\nx_i(t) \in (0,1]$ $\forall i,j \in [n]$ and $t \in \mathbb{N}$. 
\end{lemma}
The proof is included in the appendix, together with all the other omitted proofs.

\begin{lemma} \label{lem:potential_function}
	For each $i,j \in [n]$ with $\nv_{i,j} \cdot \nv_{j,i} > 0$, consider  the function $f_{i,j}^* : \mathbb{N}\to \mathbb{R}$ given by 
	\begin{align}f_{i,j}^*(t) = \left[ \nx_i(t) \cdot  \nx_j(t) \right]^{\epsilon_i \epsilon_j + e^{-t}}   \ny_{i,j}(t)^{\epsilon_j + e^{-t}}  \ny_{j,i}(t)^{\epsilon_i + e^{-t}}\,. \notag 
	\end{align}
	Then for all $t \in \mathbb{N}$, we have 
	\begin{align} \label{eq:basic_inequality_f_ij_star}
		f_{i,j}^*(t+1) \geq \Bigl(\nv_{i,j} \cdot \nv_{j,i}\Bigr)^{\epsilon_i  \epsilon_j + e^{-t}} \cdot  f_{i,j}^*(t) \,.
	\end{align}
\end{lemma}
The idea for finding the function $f_{i,j}^*(t)$ is based on the analysis of the non-damped dynamic, which showed that  in the long term players trade along cycles of length $1$ or $2$. This suggests that in the damped dynamic, such cycles are still relevant. Thus to understand the long term fortunes of the players in the damped dynamic, we look for a potential function defined on cycles $(i,j)$, where $i,j \in [n]$.

\smallskip 

Towards this end, define $f_{i,j} : \mathbb{N} \to \mathbb{R}$ as  $${f}_{i,j}(t) = \nx_i(t)^{\alpha_{i,j}(t)}   \nx_j(t)^{\alpha_{j,i}(t)}   \ny_{i,j}(t)^{\beta_{i,j}(t)}  \ny_{j,i}(t)^{\beta_{j,i}(t)},$$ where $\alpha_{i,j}, \alpha_{j,i}, \beta_{i,j}, \beta_{j,i} : \mathbb{N} \to \mathbb{R}$ are  functions that have to be determined. 

\smallskip 

Then we identify a set of constraints to ensure that ${f}_{i,j}(t+1) \geq \nv_{i,j}^{\alpha_{i,j}(t)} \cdot \nv_{j,i}^{\alpha_{j,i}(t)} \cdot {f}_{i,j}(t)\,.$
Solving the constraints shows that a good choice is obtained by setting  
\begin{align} 
	\alpha_{i,j}(t) & = \alpha_{j,i}(t) = \epsilon_i \epsilon_j + e^{-t}; \notag \\
	\beta_{i,j}(t) & = \epsilon_j + e^{-t}; \beta_{j,i}(t) = \epsilon_i + e^{-t}, \notag 
\end{align}
yielding the function $f_{i,j}^*$ from the lemma statement.

\begin{figure}[h!]
	\centering
	\subfigure[$f_{1,2}^*(t)$.]
	{
		\centering
		\includegraphics[scale = 0.46]{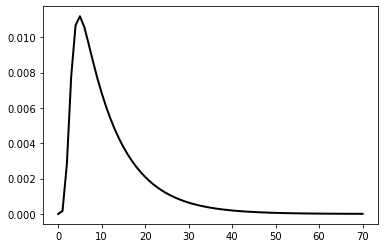}
		\label{fig:f_12}
	}
	\subfigure[$f_{1,2}^*(t) - (\nv_{1,2} \cdot \nv_{2,1})^{\epsilon_1 \epsilon_2 + e^{-t}} \cdot f_{1,2}^*(t-1)$.]
	{
		\includegraphics[scale = 0.46]{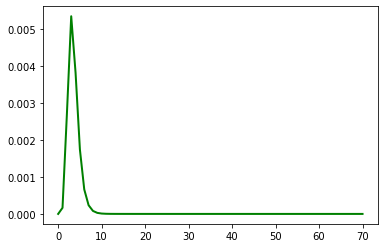}
		\label{fig:delta_12}
	}
	\caption{Illustration associated with   market for two players, where  the matrix is $\vec{v} = [[0.75, 0.61], [1, 0.75]]$, the initial quantities are $\vec{\dmx}(0) = [0.01, 0.01]$, and the initial fractions are $\vec{\dmy}(0) = [[0.9, 0.1], [0.3, 0.7]]$. Let $\epsilon_1 = 0.6$ and $\epsilon_2 = 0.4$. The matrix $\vec{v}$ is already normalized, so $v_{i,j} = \nv_{i,j}$ for all $i,j$. Figure (a) illustrates the function $f_{1,2}^*(t)$ while figure (b) illustrates the function $f_{1,2}^*(t) -  $ $(\nv_{1,2} \cdot \nv_{2,1})^{\epsilon_1 \epsilon_2 + e^{-t}} \cdot f_{1,2}^*(t-1)$, which was shown to be non-negative in Lemma~\ref{lem:potential_function}. The X axis shows the time.}
	\label{fig:function_f*}
\end{figure}

The functions $f_{i,j}^*(t)$ and   $ f_{i,j}^*(t) - (\nv_{i,j} \cdot \nv_{j,i})^{\epsilon_i \epsilon_j + e^{-t}} \cdot f_{i,j}^*(t-1)$ are illustrated in Figure~\ref{fig:function_f*}.

\begin{lemma} \label{lem:unroll_potential}
	Let $i,j \in [n]$ with $\nv_{i,j} \cdot \nv_{j,i} > 0$.
	The function  $f_{i,j}^* : \mathbb{N}\to \mathbb{R}$ from Lemma \ref{lem:potential_function} has the property: 
	$f_{i,j}^*(t) \geq \left(\nv_{i,j} \cdot \nv_{j,i}\right)^{\left(t  \epsilon_i \epsilon_j + \frac{e - e^{-t}}{e-1}\right)} \cdot f_{i,j}^*(0)\,. $
\end{lemma}
The proof of Lemma~\ref{lem:unroll_potential} follows by unrolling inequality \eqref{eq:basic_inequality_f_ij_star} from Lemma~\ref{lem:potential_function}.
Then we can lower bound the product of the amounts along cycles of length $2$ in the normalized damped system as follows.

\begin{lemma} \label{lem:bound_on_product_amounts}
	For each $i,j \in [n]$ with  $\nv_{i,j} \cdot \nv_{j,i} > 0$ and $t \in \mathbb{N}_+$ we have 
	\begin{align} \label{eq:ineq_xi_xj_product}
		& \nx_i(t) \cdot \nx_j(t) \cdot \ny_{i,j}(t)^{\frac{\epsilon_j + e^{-t}}{\epsilon_i \epsilon_j + e^{-t}}} \cdot \ny_{j,i}(t)^{\frac{\epsilon_i + e^{-t}}{\epsilon_i \epsilon_j + e^{-t}}} \geq c_{i,j} \cdot \left(\nv_{i,j} \cdot \nv_{j,i}\right)^{t + \frac{\frac{e - e^{-t}}{e-1} - t e^{-t}}{\epsilon_i \epsilon_j + e^{-t}}}   
	\end{align}
	where $c_{i,j} =  {f_{i,j}^*(0)}^{\frac{1}{\epsilon_i \epsilon_j}}$ depends on   $\vec{v}$,   $\vec{\epsilon}$, and the initial configuration.
\end{lemma}

We obtain the following lower bound on the rate of growth for the fortune of player $i$ in the normalized damped system. 

\begin{lemma} \label{prop:average_two_rounds_lb}
	For each  $i,j \in [n]$ we have $
	\frac{1}{2}\left(\dmx_i(t) + \dmx_i(t+1)\right) \geq \widetilde{b}_{i,j} \cdot \left(v_{i,j} v_{j,i}\right)^{\frac{t}{2}}\,, 
	$
	where $\widetilde{b}_{i,j}$ 
	is a constant that depends on $\vec{v}$, $\epsilon$, and the initial configuration. 
\end{lemma}

By combining the previous lemmas, we get the next theorem. 
\begin{theorem} \label{thm:main_damped}
	Let $i \in [n]$ be any player. There exists a constant  $\widetilde{c}_i > 0$ that depends on the initial configuration, the matrix $\vec{v}$, and the vector $\epsilon$, such that: 
	\begin{itemize} 
		\item $\dmx_i(t) \geq  \widetilde{c}_{i} \cdot \left({v_{i,i}}\right)^{t}$  $\forall t \in \mathbb{N}$. 
		\item $\frac{1}{2} \Bigl(\dmx_i(t) + \dmx_i(t+1)\Bigr) \geq \widetilde{c}_{i} \cdot    \max_{j \in [n]} \left({v_{i,j}} \cdot {v_{j,i}}\right)^{\frac{t}{2}}$  $\forall t \in \mathbb{N}$.
	\end{itemize}
\end{theorem}

Thus the fortune of player $i$ grows in the limit if the player has a good self loop (i.e. $v_{i,i} > 1$) or  a good cycle of length two with another player $j$ (i.e. $v_{i,j} v_{j,i} > 1$).

\section{Discussion}  \label{sec:discussion}

{\paragraph{Modeling organizational partnerships.} The tit-for-tat dynamics in networked production markets can offer a lens through which to view the structuring of organizational partnerships. This model suggests that fostering long-term growth of an organization hinges on creating relationships where reciprocal exchanges between the agents in the organization are paramount. 
For example, agents might form partnerships where resources and expertise are shared, leading to a synergy that enhances their productivity. Moreover, adopting a measured response strategy to interactions with each other may lead to enhanced stability, while still ensuring steady growth.}

\paragraph{Open questions.} We conjecture that players cannot grow any faster and that the lower bound on the rate of growth given by  Theorem~\ref{thm:main_damped}  is in fact tight up to constant factors and holds for last iterate.
\begin{conjecture}
	For each player $i \in [n]$, we have  
	$\dmx_i(t) \in \Theta\left(\max_{j \in [n]}\left(v_{i,j} \cdot v_{j,i}\right)^{\frac{t}{2}}\right)\,.$ 
\end{conjecture} 

It would also be interesting to understand if damped  updates are formally equivalent to asynchronous updates. In the latter, in each round, a player selected uniformly at random updates their strategy, while everyone else keeps theirs as before. 

Another direction is to consider a generalization of the tit-for-tat dynamic, where players use  multiplicative weight updates or other learning algorithms to update their fractions. Would such a rule  enable the players to exploit longer cycles in the graph and thus cooperate more efficiently while taking local decisions and without using monetary budgets?

\section{Acknowledgements}

Research  supported by NSF CAREER grant  CCF-2238372. The work was done in part while the author was visiting the Simons Institute for the Theory of
Computing.

\appendix 

\newpage
\section{Damped Tit-for-Tat} \label{app:damped}

In this section we include the remaining proofs from  Section~\ref{sec:damped_tit_for_tat}, analyzing the damped tit-for-tat dynamics. 


Recall the notation for the damped dynamic of Definition~\ref{def:tft_savings_lazy}. Player $i$ has $\widetilde{x}_i(t)$ units of good $i$ at time $t$. The fraction of good $i$ that player $i$ gives to player $j$ at time $t$ is $\widetilde{y}_{i,j}(t)$. 
Also recall that for the analysis, we work with a normalized version of the market, where the amounts are rescaled so that the initial amounts sum up to $1$ and the weights $v_{i,j}$ are normalized by dividing them by the largest value; see equation \eqref{eq:rescaling_damped}.

\medskip 

\noindent \textbf{Lemma~\ref{lem:normalized_system_relation} (restated).} \emph{Suppose 
	\begin{itemize}
		\item $(\vec{\dmx}(t), \vec{\dmy}(t))_{t \geq 0}$ are the variables of the dynamical system of Definition~\ref{def:tft_savings_lazy} with matrix $\vec{v}$, vector $\vec{\epsilon}$, initial amounts $\vec{\dmx}(0)$, and initial fractions $\vec{\dmy}(0)$.
		\item $(\vec{\nx}(t), \ny(t))_{t\geq 0}$ are the variables of the dynamical system of Definition~\ref{def:tft_savings_lazy} with matrix $\vec{\omega}$, vector $\vec{\epsilon}$, initial amounts $\vec{\nx}(0)$, and initial fractions $\vec{\ny}(0)$ as defined in (\ref{eq:rescaling_damped}). 
	\end{itemize} 
	Then $\nx_i(t) = \frac{\dmx_i(t)}{\sigma \delta^{t}}$, $\ny_{i,j}(t) = \dmy_{i,j}(t)$, $\nv_{i,j} \in [0,1]$, and $\nx_i(t) \in (0,1]$ $\forall i,j \in [n]$ and $t \in \mathbb{N}$.} \\
\begin{proof}
	By definition, $\ny_{i,j}(0) = \vec{\dmy}(0)$ and $\nx_{i}(0) = \frac{\dmx_i(0)}{\sigma \cdot \delta^0}$  $\forall i,j \in [n]$. We prove the identities in the lemma statement by induction: assume they hold at time $t$ and  prove they also hold at  $t+1$. The induction step will be proved in two parts, (a) and (b):

	\noindent (a). For each player $i$, we have:  
	\begin{align}
		\nx_{i}(t+1) & = \sum_{k=1}^n \nv_{i,k} \cdot \ny_{k,i}(t) \cdot \nx_k(t)   = \sum_{k=1}^n \left(\frac{v_{i,k}}{\delta}\right) \cdot \dmy_{k,i}(t) \cdot \left( \frac{\dmx_k(t)}{\sigma \delta^t} \right)   = \frac{\dmx_i(t+1)}{\sigma \delta^{t+1}}\,. \notag 
	\end{align} 
	
	\noindent (b). Using the induction hypothesis and part (a), for each $i,j \in [n]$ we have \begin{align} 
		\ny_{i,j}(t+1) & \; = \epsilon_i \cdot \frac{\nv_{i,j} \cdot \ny_{j,i}(t) \cdot \nx_j(t)}{\nx_i(t+1)} + (1-\epsilon_i ) \ny_{i,j}(t) \notag \\ 
		& \; = \epsilon_i \cdot \frac{\left(\frac{v_{i,j}}{\delta}\right) \cdot \dmy_{j,i}(t) \cdot \left(\frac{\dmx_j(t)}{\sigma \delta^{t}}\right)}{\left(\frac{\dmx_i(t+1)}{\sigma \delta^{t+1}}\right)} + (1-\epsilon_i ) \dmy_{i,j}(t)  \notag \\
		& \; = \epsilon_i \cdot \frac{v_{i,j} \cdot \dmy_{j,i}(t) \cdot \dmx_j(t)}{\dmx_i(t+1)} + (1-\epsilon_i ) \dmy_{i,j}(t) = \dmy_{i,j}(t+1)\,. \notag
	\end{align}
	This completes the induction step. 
	
	Finally, to see that $\nx_i(t) \in (0, 1]$ for all $i \in [n]$ and $t \in \mathbb{N}$, observe that the maximum amount that a player can have at time $t$ is taking the sum of amounts at time zero and routing them through the cycle with the highest geometric mean (of the weighted directed graph with  weights $\nv$) repeatedly. However, by the definition in (\ref{eq:rescaling_damped}), we have $\sum_{i=1}^{n}\nx_i(0) = 1$ and all $\nv_{i,j}\leq 1$, thus the amount of each good $i$ will  be upper bounded by $1$ throughout time.
\end{proof}

\noindent \textbf{Lemma~\ref{lem:potential_function} (restated).}
\emph{For each $i,j \in [n]$ with $\nv_{i,j} \cdot \nv_{j,i} > 0$, consider  the function $f_{i,j}^* : \mathbb{N}\to \mathbb{R}$ given by }
\begin{align}f_{i,j}^*(t) = \left[ \nx_i(t) \cdot  \nx_j(t) \right]^{\epsilon_i \epsilon_j + e^{-t}}   \ny_{i,j}(t)^{\epsilon_j + e^{-t}}  \ny_{j,i}(t)^{\epsilon_i + e^{-t}} \,. \notag 
\end{align}
Then for all $t \in \mathbb{N}$, we have:  
$f_{i,j}^*(t+1) \geq \Bigl(\nv_{i,j} \cdot \nv_{j,i}\Bigr)^{\epsilon_i  \epsilon_j + e^{-t}} \cdot  f_{i,j}^*(t) \,. $
\begin{proof}
	We first define a function ${f}_{i,j} : \mathbb{N} \to \mathbb{R}$ in symbolic form:
	$$
	{f}_{i,j}(t) = \nx_i(t)^{\alpha_{i,j}(t)}  \cdot \nx_j(t)^{\alpha_{j,i}(t)}  \cdot \ny_{i,j}(t)^{\beta_{i,j}(t)} \cdot \ny_{j,i}(t)^{\beta_{j,i}(t)}
	$$
	where $\alpha_{i,j}, \alpha_{j,i}, \beta_{i,j}, \beta_{j,i} : \mathbb{N}\to \mathbb{R}_{+}$ will be set later. Then we will set constraints to ensure the next inequality holds: ${f}_{i,j}(t+1) \geq \nv_{i,j}^{\alpha_{i,j}(t)} \cdot \nv_{j,i}^{\alpha_{j,i}(t)} \cdot {f}_{i,j}(t) \; \; \; \; \forall t \in \mathbb{N}\,.
	$
	Let $k,\ell \in [n]$ be arbitrary such that $\nv_{k,\ell} \cdot v_{\ell, k} > 0$.
	Applying the weighted AM-GM inequality in the definition of the update for $\ny_{k,\ell}(t+1)$, we obtain:
	\begin{align} 
		\ny_{k,\ell}(t+1)&  =  \epsilon_k \cdot  \frac{\nv_{k,\ell} \cdot \ny_{\ell,k}(t) \cdot \nx_{\ell}(t)}{\nx_k(t+1)} + (1-\epsilon_k) \ny_{k,\ell}(t) \notag \\
  & \geq \left(\frac{\nv_{k,\ell} \cdot \ny_{\ell,k}(t) \cdot \nx_{\ell}(t)}{\nx_{k}(t+1)}\right)^{\epsilon_k}  \ny_{k,\ell}(t)^{1-\epsilon_k}  \label{eq:y_ij_amgm} 
	\end{align} 
	
	Using (\ref{eq:y_ij_amgm}) we get a  lower bound on $f_{i,j}(t+1)$:  
	\begin{align} \label{eq:f_ij_lb_1}
		f_{i,j}(t+1) & = \nx_i(t+1)^{\alpha_{i,j}(t+1)}   \nx_j(t+1)^{\alpha_{j,i}(t+1)}   \ny_{i,j}(t+1)^{\beta_{i,j}(t+1)}  \ny_{j,i}(t+1)^{\beta_{j,i}(t+1)} \notag \\
		& \geq \nx_i(t+1)^{\alpha_{i,j}(t+1)}   \nx_j(t+1)^{\alpha_{j,i}(t+1)} \notag \\
  & {\hspace{10pt}} \cdot   \left(\frac{\nv_{i,j}  \ny_{j,i}(t)  \nx_{j}(t)}{\nx_{i}(t+1)}\right)^{\epsilon_i \cdot \beta_{i,j}(t+1)} \ny_{i,j}(t)^{(1-\epsilon_i) \beta_{i,j}(t+1)} \notag \\
		& {\hspace{10pt}} \cdot \left(\frac{\nv_{j,i} \cdot \ny_{i,j}(t) \cdot \nx_{i}(t)}{\nx_{j}(t+1)}\right)^{\epsilon_j \cdot \beta_{j,i}(t+1)}   \ny_{j,i}(t)^{(1-\epsilon_j) \beta_{j,i}(t+1)} \notag \\
		& = \nx_i(t+1)^{\alpha_{i,j}(t+1)-\epsilon_i \cdot \beta_{i,j}(t+1)}  \cdot \nx_j(t+1)^{\alpha_{j,i}(t+1)-\epsilon_j \cdot \beta_{j,i}(t+1)} \notag \\
  &  {\hspace{10pt}}  \cdot \Bigl(\nx_{i}(t)\Bigr)^{\epsilon_j \cdot \beta_{j,i}(t+1)} \cdot  \Bigl(\nx_{j}(t)\Bigr)^{\epsilon_i \cdot \beta_{i,j}(t+1)} \notag \\
		& {\hspace{10pt}} \cdot \nv_{i,j}^{\epsilon_i \cdot \beta_{i,j}(t+1)} \cdot \nv_{j,i}^{\epsilon_j \cdot \beta_{j,i}(t+1)} \notag \\
  &  {\hspace{10pt}} \cdot 
		\ny_{i,j}(t)^{\epsilon_j \cdot \beta_{j,i}(t+1) + (1-\epsilon_i)\beta_{i,j}(t+1)}  \cdot 
		\ny_{j,i}(t)^{\epsilon_i \cdot \beta_{i,j}(t+1) + (1-\epsilon_j)\beta_{j,i}(t+1)}\,.
	\end{align}
	We will impose the next set of  constraints for all $t $ on the functions $\alpha_{i,j}, \alpha_{j,i}, \beta_{i,j}$, and $\beta_{j,i}$, in order to ensure that the exponents in the expressions above are positive:
	\begin{equation} \label{eq:constraints_1}
		\begin{cases}
			& \alpha_{i,j}(t) , \alpha_{j,i}(t),  \beta_{i,j}(t),\beta_{j,i}(t) \geq 0 \\
			& \alpha_{i,j}(t) \geq \epsilon_i \cdot \beta_{i,j}(t)   \text{  and  } \alpha_{j,i}(t) \geq \epsilon_j \cdot \beta_{j,i}(t) \,. 
		\end{cases}
	\end{equation}
	By definition of the  update rule, we have 
	\begin{align} \label{eq:simple_inequality_xixj}
		& \nx_i({t+1})  = \sum_{k =1}^n \nv_{i,k}  \ny_{k,i}(t)  \nx_k(t)  \geq \nv_{i,j}  \ny_{j,i}(t)  \nx_j(t)  \text{   and   }  
		\nx_j({t+1})   \geq \nv_{j,i}  \ny_{i,j}(t)  \nx_i(t)\,.
	\end{align}
	Combining (\ref{eq:simple_inequality_xixj}) with  the constraints in (\ref{eq:constraints_1}) and the inequalities $\nx_{k}(t) < 1$ for all $k \in [n]$, we obtain the inequalities:
	\begin{align} \label{eq:lower_bound_amounts_1}
		&  \bullet \; \; \nx_i(t+1)^{\alpha_{i,j}(t+1)-\epsilon_i\cdot \beta_{i,j}(t+1)}  \geq \Bigl(\nv_{i,j} \cdot \ny_{j,i}(t) \cdot \nx_j(t) \Bigr)^{\alpha_{i,j}(t+1)-\epsilon_i\cdot \beta_{i,j}(t+1)} \;  \notag  \\
		&  \bullet \; \;  \nx_j(t+1)^{\alpha_{j,i}(t+1)-\epsilon_j\cdot \beta_{j,i}(t+1)}   \geq \Bigl( \nv_{j,i} \cdot \ny_{i,j}(t) \cdot \nx_i(t) \Bigr)^{\alpha_{j,i}(t+1)-\epsilon_j\cdot \beta_{j,i}(t+1)}\,. 
	\end{align}
	Using  (\ref{eq:lower_bound_amounts_1}) in (\ref{eq:f_ij_lb_1}) we can further bound $f_{i,j}(t+1)$ as follows:
	\begin{align} \label{eq:f_ij_lb_2}
		f_{i,j}(t+1) & \geq \Bigl(\nv_{i,j} \cdot \ny_{j,i}(t) \cdot \nx_j(t) \Bigr)^{\alpha_{i,j}(t+1)-\epsilon_i\cdot \beta_{i,j}(t+1)}  \cdot  \Bigl( \nv_{j,i} \cdot \ny_{i,j}(t) \cdot \nx_i(t) \Bigr)^{\alpha_{j,i}(t+1)-\epsilon_j\cdot \beta_{j,i}(t+1)} \notag \\
		&{\hspace{10pt}} \cdot \Bigl(\nx_{i}(t)\Bigr)^{\epsilon_j \cdot \beta_{j,i}(t+1)} \cdot \Bigl(\nx_{j}(t)\Bigr)^{\epsilon_i \cdot \beta_{i,j}(t+1)}  \cdot \nv_{i,j}^{\epsilon_i \cdot \beta_{i,j}(t+1)} \cdot \nv_{j,i}^{\epsilon_j \cdot \beta_{j,i}(t+1)}   \notag \\
		&{\hspace{10pt}} 
		\cdot \ny_{i,j}(t)^{\epsilon_j \cdot \beta_{j,i}(t+1) + (1-\epsilon_i)\beta_{i,j}(t+1)}  \cdot 
		\ny_{j,i}(t)^{\epsilon_i \cdot \beta_{i,j}(t+1) + (1-\epsilon_j)\beta_{j,i}(t+1)} \notag \\
		& = \Bigl( \nv_{i,j}^{\alpha_{i,j}(t+1)} \cdot \nv_{j,i}^{\alpha_{j,i}(t+1)}\Bigr)   \cdot \Bigl(\nx_{i}(t)^{\alpha_{j,i}(t+1)} \cdot \nx_{j}(t)^{\alpha_{i,j}(t+1)} \Bigr) \notag \\
		&{\hspace{10pt}} \cdot \Bigl(\ny_{i,j}(t)\Bigr)^{\alpha_{j,i}(t+1)+(1-\epsilon_i)\cdot\beta_{i,j}(t+1)} \cdot \Bigl(\ny_{j,i}(t)\Bigr)^{\alpha_{i,j}(t+1)+(1-\epsilon_j)\cdot\beta_{j,i}(t+1)} \,.
	\end{align}
	We would like the  last expression in (\ref{eq:f_ij_lb_2}) to be at least  $\left(\nv_{i,j}^{\alpha_{i,j}(t)} \cdot \nv_{j,i}^{\alpha_{j,i}(t)}\right) \cdot f(t)$. Thus we  require the following inequality to hold:
	\begin{align} \label{eq:ideal_lb}
		&	 v_{i,j}^{\alpha_{i,j}(t+1)} \cdot v_{j,i}^{\alpha_{j,i}(t+1)} \cdot x_{i}(t)^{\alpha_{j,i}(t+1)} \cdot x_{j}(t)^{\alpha_{i,j}(t+1)}  \notag \\
		& {\hspace{39pt}}
		\cdot		y_{i,j}(t)^{\alpha_{j,i}(t+1)+(1-\epsilon_i) \beta_{i,j}(t+1)}  \notag \\
		& {\hspace{39pt}} \cdot y_{j,i}(t)^{\alpha_{i,j}(t+1)+(1-\epsilon_j)\beta_{j,i}(t+1)} \notag \\
		&\geq  v_{i,j}^{\alpha_{i,j}(t)} \cdot v_{j,i}^{\alpha_{j,i}(t)}   \cdot x_{i}(t)^{\alpha_{i,j}(t)} \cdot x_{j}(t)^{\alpha_{j,i}(t)}  \cdot y_{i,j}(t)^{\beta_{i,j}(t)} \cdot y_{j,i}(t)^{\beta_{j,i}(t)}   \,.
	\end{align}
	Since $\nx_i(t), y_{i,j}(t) \leq 1 \leq 1$ for all $i,j \in [n] $ and $t \in \mathbb{N}$, to ensure that inequality (\ref{eq:ideal_lb})  holds, we will require that the exponents on the left hand side are smaller than those on the right hand side. 
	Putting  this together with the inequalities in (\ref{eq:constraints_1}) gives the next constraints for all $t \in \mathbb{N}$:
	\begin{enumerate}[(a)]
		\item $	\alpha_{i, j}(t), \alpha_{j,i}(t),  \beta_{i,j}(t), \beta_{j,i}(t) \geq 0$ 
		\item $\alpha_{i,j}(t) \geq  \max\{ \alpha_{i,j}(t+1), \alpha_{j,i}(t+1)\}$ and $\alpha_{j,i}(t) \geq \max \{ \alpha_{j,i}(t+1), \alpha_{i,j}(t+1)\}$ 
		\item $\alpha_{i,j}(t) \geq \epsilon_i \cdot \beta_{i,j}(t)$ 
		and  $\alpha_{j,i}(t) \geq \epsilon_j \cdot \beta_{j,i}(t)$ 
		\item $\alpha_{j,i}(t+1) + (1-\epsilon_i) \cdot \beta_{i,j}(t+1) \leq \beta_{i,j}(t)$ and  $\alpha_{i,j}(t+1) + (1-\epsilon_j) \cdot \beta_{j,i}(t+1) \leq \beta_{j,i}(t)$ 
	\end{enumerate}
	For each $ t \in \mathbb{N}$, set 
	\begin{align} \label{eq:setting_variables}
		& \alpha_{i,j}(t) = \alpha_{j,i}(t) = \epsilon_i \cdot \epsilon_j + e^{-t} ; \; \; \; 	\beta_{i,j}(t) = \epsilon_j + e^{-t}   ; \; \; \; 	\beta_{j,i}(t) = \epsilon_i + e^{-t} \,.
	\end{align}
	We check that  the constraints in (a-d) are met for the  choice of variables in \eqref{eq:setting_variables}.
	
	Clearly (a)  holds. 
	The function $\epsilon_i \cdot \epsilon_j + e^{-t}$ is monotonically decreasing and $\alpha_{i,j}(t) = \alpha_{j,i}(t)$, so  (b) holds.
	
	For the first constraint in (c), we require
$\alpha_{i,j}(t) = \epsilon_i \cdot \epsilon_j + e^{-t} \geq \epsilon_i \cdot (\epsilon_j + e^{-t}),$
	or equivalently $e^{-t}  (1 - \epsilon_i) \geq 0$. The latter is clearly true since $\epsilon_i \in (0, 1]$. The second constraint in (c)  holds by symmetry  since $\alpha_{i,j} = \alpha_{j,i}$. 
	
	The first inequality in (d) requires that 
	\begin{align} \label{eq:checking_g}
		& \alpha_{j,i}(t+1) + (1-\epsilon_i)  \beta_{i,j}(t+1)  = \epsilon_i  \epsilon_j + e^{-t-1} + (1-\epsilon_i) (\epsilon_j + e^{-t-1}) \leq \beta_{i,j}(t) = \epsilon_j + e^{-t}  \notag \\
		& \iff \epsilon_i \epsilon_j + e^{-t-1} + \epsilon_j + e^{-t-1} - \epsilon_i \epsilon_j - \epsilon_i \cdot e^{-t-1} \leq \epsilon_j + e^{-t} \notag \\ 
		&  \iff  2 \leq e + \epsilon_i\,.
	\end{align} 
	The last inequality in (\ref{eq:checking_g}) holds, and so  (d) holds. 
	
	Let $f_{i,j}^*$ be the instantiation of $f_{i,j}$ with the choice of exponents in (\ref{eq:setting_variables}): 
	\begin{align} 
		f_{i,j}^*(t) & = \nx_i(t)^{\epsilon_i \epsilon_j + e^{-t}}  \cdot \nx_j(t)^{\epsilon_i \epsilon_j + e^{-t}}  \cdot \ny_{i,j}(t)^{\epsilon_j + e^{-t}} \cdot \ny_{j,i}(t)^{\epsilon_i + e^{-t}}\,. \notag 
	\end{align} 
	Then we have 
	\begin{align}
		f_{i,j}^*(t+1) & \geq \nv_{i,j}^{\epsilon_i \epsilon_j + e^{-t}} \cdot \nv_{j,i}^{\epsilon_i \epsilon_j + e^{-t}}  \cdot \nx_{i}(t)^{\epsilon_i \epsilon_j + e^{-t}} \cdot \nx_j(t)^{\epsilon_i \epsilon_j + e^{-t}}   \cdot \ny_{i,j}(t)^{\epsilon_j + e^{-t}} \cdot \ny_{j,i}(t)^{\epsilon_i  + e^{-t}} \notag \\
		& = \nv_{i,j}^{\epsilon_i \epsilon_j + e^{-t}} \cdot \nv_{j,i}^{\epsilon_i \epsilon_j + e^{-t}} \cdot  f_{i,j}^*(t)\,.\notag 
	\end{align}
	This completes the proof of the lemma.
\end{proof}

\noindent \textbf{Lemma~\ref{lem:unroll_potential} (restated).} 
\emph{Let $i,j \in [n]$ with $\nv_{i,j} \cdot \nv_{j,i} > 0$.
	The function $f_{i,j}^* : \mathbb{N}\to \mathbb{R}$ from Lemma \ref{lem:potential_function} has the property: }
$f_{i,j}^*(t) \geq \left(\nv_{i,j} \cdot \nv_{j,i}\right)^{\left(t  \epsilon_i \epsilon_j + \frac{e - e^{-t}}{e-1}\right)} \cdot f_{i,j}^*(0)\,. $
\begin{proof}
	By Lemma~\ref{lem:potential_function}, the following  inequality holds for all $t \geq 1$: 
	\begin{align} \label{eq:concrete_inequality_to_unroll_f}
		f_{i,j}^*(t) & \geq \left(\nv_{i,j} \cdot \nv_{j,i}\right)^{\epsilon_i \epsilon_j + e^{-t}} \cdot f_{i,j}^*(t-1)\,.
	\end{align}
	Unrolling (\ref{eq:concrete_inequality_to_unroll_f}) by considering all rounds $\ell \in \{1, \ldots, t\}$, we get:
	\begin{align} \label{eq:bound_unroll_f}
		f_{i,j}^*(t) & \geq \left( \nv_{i,j} \cdot \nv_{j,i} \right)^{\sum_{\ell=1}^{t} \left(\epsilon_i \epsilon_j  + e^{-\ell}\right)} \cdot f_{i,j}^*(0)   \,. 
	\end{align}
	Substituting the identity 
	$\sum_{\ell=1}^{t} \left(\epsilon_i \epsilon_j + e^{-\ell}\right) = t \cdot \epsilon_i \epsilon_j + \frac{e - e^{-t}}{e-1}\;$
	in (\ref{eq:bound_unroll_f}) gives the  inequality required by the lemma statement.
\end{proof}

\noindent \textbf{Lemma~\ref{lem:bound_on_product_amounts} (restated).} 
\emph{For each $i,j \in [n]$ with  $\nv_{i,j} \cdot \nv_{j,i} > 0$ and $t \in \mathbb{N}_+$ we have}
\begin{align} \label{eq:ineq_xi_xj_product_restated}
	& \nx_i(t) \cdot \nx_j(t) \cdot \ny_{i,j}(t)^{\frac{\epsilon_j + e^{-t}}{\epsilon_i \epsilon_j + e^{-t}}} \cdot \ny_{j,i}(t)^{\frac{\epsilon_i + e^{-t}}{\epsilon_i \epsilon_j + e^{-t}}} \geq c_{i,j} \cdot \left(\nv_{i,j} \cdot \nv_{j,i}\right)^{t + \frac{\frac{e - e^{-t}}{e-1} - t e^{-t}}{\epsilon_i \epsilon_j + e^{-t}}}   
\end{align}
\emph{where $c_{i,j} =  {f_{i,j}^*(0)}^{\frac{1}{\epsilon_i \epsilon_j}}$ depends on $\vec{v}$, $\vec{\epsilon}$, and the initial configuration.}
\begin{proof}
	Lemma~\ref{lem:unroll_potential} gives: 
	$f_{i,j}^*(t)  \geq \left( \nv_{i,j} \cdot \nv_{j,i} \right)^{t \cdot \epsilon_i \epsilon_j + \frac{e - e^{-t}}{e-1}} \cdot f_{i,j}^*(0) \; ({*}).$
	Substituting the definition of $f_{i,j}^*(t)$ in ({*}), we get
	\begin{align}
		& \Bigl(\nx_i(t) \cdot \nx_j(t)\Bigr)^{\epsilon_i \cdot \epsilon_j + e^{-t}} \cdot \ny_{i,j}(t)^{\epsilon_j + e^{-t}} \cdot \ny_{j,i}(t)^{\epsilon_i + e^{-t}}  \geq f_{i,j}^*(0) \cdot \left(\nv_{i,j} \cdot \nv_{j,i}\right)^{t \cdot \epsilon_i \epsilon_j + \frac{e - e^{-t}}{e-1}}\,. 
	\end{align}
	Taking root of order $\epsilon_i \epsilon_j + e^{-t}$ on both sides and using the inequality  $f_{i,j}^*(0) \leq 1$ gives
	\begin{align} \label{eq:lower_bound_unrolled_almost_final}
		\nx_i(t) \cdot \nx_j(t) \cdot \ny_{i,j}(t)^{\frac{\epsilon_j + e^{-t}}{\epsilon_i  \epsilon_j + e^{-t}}} \cdot \ny_{j,i}(t)^{\frac{\epsilon_i + e^{-t}}{\epsilon_i  \epsilon_j + e^{-t}}}  & \geq f_{i,j}^*(0)^{\frac{1}{\epsilon_i \cdot \epsilon_j + e^{-t}}} \cdot \left(\nv_{i,j} \cdot \nv_{j,i}\right)^{\frac{t \cdot \epsilon_i \epsilon_j + \frac{e - e^{-t}}{e-1}}{\epsilon_i \cdot \epsilon_j + e^{-t}}} \notag \\
		& \geq f_{i,j}^*(0)^{\frac{1}{\epsilon_i \epsilon_j}} \cdot \left(\nv_{i,j} \cdot \nv_{j,i}\right)^{\frac{t \cdot \epsilon_i \epsilon_j + \frac{e - e^{-t}}{e-1}}{\epsilon_i \cdot \epsilon_j + e^{-t}}}\,.
	\end{align}
	Let $c_{i,j} = {f_{i,j}^*(0)}^{\frac{1}{\epsilon_i \epsilon_j}}$. Replacing  
	\begin{align}
		& {\left(t  \epsilon_i \epsilon_j + \frac{e - e^{-t}}{e-1}\right)} \cdot \frac{1}{{\epsilon_i  \epsilon_j + e^{-t}}} = t + {\left(\frac{e - e^{-t}}{e-1} - t e^{-t}\right)}\frac{1}{\epsilon_i \epsilon_j + e^{-t}} \notag 
	\end{align}
	in  (\ref{eq:lower_bound_unrolled_almost_final}) gives the  inequality required by the lemma. 
\end{proof}

\noindent \textbf{Lemma~\ref{prop:average_two_rounds_lb} (restated).}
\emph{For each  $i,j \in [n]$ we have $
	\frac{1}{2}\left(\dmx_i(t) + \dmx_i(t+1)\right) \geq \widetilde{b}_{i,j} \cdot \left(v_{i,j} v_{j,i}\right)^{\frac{t}{2}}\,, 
	$
	where $\widetilde{b}_{i,j}$ 
	is a constant that depends on $\vec{v}$, $\epsilon$, and the initial configuration. 
}
\begin{proof}
	Inequality~\ref{eq:simple_inequality_xixj} gives
 \begin{align} 
\nx_{i}(t+1)  \geq \nv_{i,j} \cdot \ny_{j,i}(t) \cdot \nx_j(t) \; \; \; ({**}) \,. \notag 
 \end{align} 
	Lemma~\ref{lem:bound_on_product_amounts} guarantees   
	\begin{align}  \label{eq:bound_on_product_amounts_reversed}
		&  c_{i,j} \cdot \left(\nv_{i,j} \cdot \nv_{j,i}\right)^{t + \frac{\frac{e - e^{-t}}{e-1} - t e^{-t}}{\epsilon_i \epsilon_j + e^{-t}}}  \leq 
		\nx_i(t) \cdot \nx_j(t) \cdot \ny_{i,j}(t)^{\frac{\epsilon_j + e^{-t}}{\epsilon_i \epsilon_j + e^{-t}}} \cdot \ny_{j,i}(t)^{\frac{\epsilon_i + e^{-t}}{\epsilon_i \epsilon_j + e^{-t}}}
	\end{align}
	where $c_{i,j} =  {f_{i,j}^*(0)}^{\frac{1}{\epsilon_i \epsilon_j}}$. 
	
	Using $(**)$ in inequality  (\ref{eq:bound_on_product_amounts_reversed}), we obtain
	\begin{align} \label{eq:simple_bounds_average_consecutive_xi}
		c_{i,j} \cdot \left(\nv_{i,j} \cdot \nv_{j,i}\right)^{t + \frac{\frac{e - e^{-t}}{e-1} - t e^{-t}}{\epsilon_i \epsilon_j + e^{-t}}}  & \leq 
		\nx_i(t) \cdot \frac{\nv_{i,j} \cdot \ny_{j,i}(t) \cdot \nx_j(t)}{\nv_{i,j}}  \cdot \ny_{i,j}(t)^{\frac{\epsilon_j + e^{-t}}{\epsilon_i \epsilon_j + e^{-t}}} \cdot \ny_{j,i}(t)^{\frac{\epsilon_i + e^{-t}}{\epsilon_i \epsilon_j + e^{-t}}-1} \notag \\
		& \leq \nx_i(t) \cdot \frac{\nx_i(t+1)}{\nv_{i,j}}  \cdot  \ny_{i,j}(t)^{\frac{\epsilon_j + e^{-t}}{\epsilon_i \epsilon_j + e^{-t}}} \cdot \ny_{j,i}(t)^{\frac{\epsilon_i + e^{-t}}{\epsilon_i \epsilon_j + e^{-t}}-1}  \,.
	\end{align}
	Recall $\ny_{i,j}(t), \ny_{j,i}(t) \in (0,1)$. Taking square root on both sides of (\ref{eq:simple_bounds_average_consecutive_xi})  and using the AM-GM inequality, we can bound  the average amount of player $i$ over any two consecutive rounds:
	\begin{align}
		\frac{\nx_i(t) + \nx_i(t+1)}{2} & \geq \sqrt{\nx_i(t) \cdot \nx_i(t+1)} \notag \\
		& \geq \sqrt{c_{i,j} \cdot \nv_{i,j}}  \cdot \left(\nv_{i,j} \cdot \nv_{j,i}\right)^{\frac{t}{2} + \frac{\frac{e - e^{-t}}{e-1} - t e^{-t}}{2\epsilon_i \epsilon_j + 2e^{-t}}} \cdot \ny_{i,j}(t)^{-\frac{\epsilon_j + e^{-t}}{2\epsilon_i \epsilon_j + 2e^{-t}}} \cdot \ny_{j,i}(t)^{\frac{1}{2}-\frac{\epsilon_i + e^{-t}}{2\epsilon_i \epsilon_j + 2e^{-t}}}   \notag \\
		& \geq \sqrt{c_{i,j} \cdot \nv_{i,j}} \cdot \left(\nv_{i,j} \cdot \nv_{j,i}\right)^{\frac{t}{2} + \frac{\frac{e - e^{-t}}{e-1} - t e^{-t}}{2\epsilon_i \epsilon_j + 2e^{-t}}}  \notag \\
		& \geq \sqrt{c_{i,j} \cdot \nv_{i,j}} \cdot \left(\nv_{i,j} \cdot \nv_{j,i}\right)^{\frac{t}{2} + \frac{e}{2\epsilon_i \epsilon_j (e-1)}}  = \sqrt{c_{i,j}} \cdot \nv_{i,j}^{\frac{t+1}{2} +   \frac{e}{2\epsilon_i \epsilon_j (e-1)}} \cdot \nv_{j,i}^{\frac{t}{2} + \frac{e}{2\epsilon_i \epsilon_j (e-1)}}\,. \notag
	\end{align} 
	
	Setting $\gamma = \frac{e}{2\epsilon_i \epsilon_j (e-1)}$, we obtain that for all $t \in \mathbb{N}$: 
	\begin{align}  \label{eq:inequality_average_amounts_normalized}
		\frac{\nx_i(t) + \nx_i(t+1)}{2} \geq \sqrt{c_{i,j}} \cdot \bigl(\nv_{i,j}\bigr)^{\frac{t+1}{2}+\gamma} \cdot \bigl(\nv_{j,i}\bigr)^{\frac{t}{2}+\gamma} \,.
	\end{align} 
	
	From the definition of the normalized system in (\ref{eq:rescaling_damped}), we have $\omega_{i,j} = v_{i,j}/\delta$, for $\delta = \max_{k,\ell} v_{k,\ell}$. 
		By Lemma~\ref{lem:normalized_system_relation}, we have   
	\begin{align}  \label{eq:two_identities_normalized}
		\nx_i(t) = \frac{\dmx_i(t)}{\sigma \delta^{t}} \text{ and } \ny_{i,j}(t) = \dmy_{i,j}(t) \; \; \;  \forall i,j \in [n] \,.
	\end{align} 
	Substituting the identities from \eqref{eq:two_identities_normalized} in inequality (\ref{eq:inequality_average_amounts_normalized}), we get that for all $ t \in \mathbb{N}$:
	
	\begin{align} \label{eq:average_amounts_initial_averagetwo}
		\frac{\dmx_i(t)}{2 \sigma \delta^{t}} + \frac{\dmx_i(t+1)}{2 \sigma \delta^{t+1}} & \geq \sqrt{c_{i,j}}  \left(\frac{v_{i,j}}{\delta}\right)^{\frac{t+1}{2}+\gamma}  \left(\frac{v_{j,i}}{\delta}\right)^{\frac{t}{2}+\gamma} \iff \notag \\
		\frac{1}{2}\left({\dmx_i(t)} + {\dmx_i(t+1)}\right) & \geq   \left(\frac{\sigma \cdot \sqrt{c_{i,j}} \cdot \left(v_{i,j}\right)^{\frac{1}{2} + \gamma} \cdot \left(v_{j,i}\right)^{\gamma}}{\delta^{2\gamma+1/2}} \right)   {v_{i,j}}^{\frac{t}{2}}  {v_{j,i}}^{\frac{t}{2}}  \,.
	\end{align}
	
	Define 
	\begin{align} 
		\widetilde{b}_{i,j} = \frac{\sigma \cdot \sqrt{c_{i,j}}  \left(v_{i,j}\right)^{\frac{1}{2} + \gamma} \left(  v_{j,i}\right)^{\gamma}}{\delta^{2\gamma+1/2}} \,,  \notag 
	\end{align}
	where 
	\begin{align} 
		c_{i,j} & = f_{i,j}^*(0)^{\frac{1}{\epsilon_i\epsilon_j}} \; ; \;  
		\gamma  = \frac{e}{2\epsilon_i \epsilon_j (e-1)} \; ; 
		\sigma = \sum_{k=1}^n x_k(0) \; ;  \; 
		\delta  = \max_{k,\ell} v_{k,\ell} \,. \notag  
	\end{align}
	Then substituting $\widetilde{b}_{i,j}$ in the last inequality of (\ref{eq:average_amounts_initial_averagetwo}) yields the lemma statement.
\end{proof}

Combining the previous lemmas, we get the next theorem.\\  

\noindent \textbf{Theorem~\ref{thm:main_damped} (restated).}
\emph{Let $i \in [n]$ be any player. There exists a constant  $\widetilde{c}_i > 0$ that depends on the initial configuration, the matrix $\vec{v}$, and the vector $\epsilon$, such that:} 
\begin{itemize} 
	\item $\dmx_i(t) \geq  \widetilde{c}_{i} \cdot \left({v_{i,i}}\right)^{t}$  $\forall t \in \mathbb{N}$. 
	\item $\frac{1}{2} \Bigl(\dmx_i(t) + \dmx_i(t+1)\Bigr) \geq \widetilde{c}_{i} \cdot    \max_{j \in [n]} \left({v_{i,j}} \cdot {v_{j,i}}\right)^{\frac{t}{2}}$  $\forall t \in \mathbb{N}$.
\end{itemize}
\emph{Thus the average amount taken the last two rounds of a player $i$  grows at a rate greater than or equal to the geometric mean of the best cycle of length at most two that $i$  is part of.
	If player $i$  has a ``good'' self loop, with $v_{i,i} > 0$, then its amount also grows at a rate greater than or equal to $v_{i,i}$. }
\begin{proof}
	For the first inequality in the theorem statement, recall Lemma~\ref{lem:bound_on_product_amounts} gives 
	\begin{align}  \label{eq:inequality_lost_number}
		& \nx_i(t) \cdot \nx_j(t) \cdot \ny_{i,j}(t)^{\frac{\epsilon_j + e^{-t}}{\epsilon_i \epsilon_j + e^{-t}}} \cdot \ny_{j,i}(t)^{\frac{\epsilon_i + e^{-t}}{\epsilon_i \epsilon_j + e^{-t}}}  \geq c_{i,j} \cdot \left(\nv_{i,j} \cdot \nv_{j,i}\right)^{t + \frac{\frac{e - e^{-t}}{e-1} - t e^{-t}}{\epsilon_i \epsilon_j + e^{-t}}} \;  \forall i,j \in [n] \; \forall t \in \mathbb{N}\,. 
	\end{align}
	where $c_{i,j} =  {f_{i,j}^*(0)}^{\frac{1}{\epsilon_i \epsilon_j}}$.
	Setting $i=j$ and taking root on both sides of    \eqref{eq:inequality_lost_number}, we get:  
	\begin{align}  \label{eq:lb_nxi_alone}
		\nx_i(t) \cdot \ny_{i,i}(t)^{\frac{\epsilon_i + e^{-t}}{\epsilon_i \epsilon_i + e^{-t}}}  & \geq \sqrt{c_{i,i}} \cdot \left(\nv_{i,i}\right)^{t + \frac{\frac{e - e^{-t}}{e-1} - t e^{-t}}{\epsilon_i \epsilon_i + e^{-t}}}  \notag \\
  & \geq \sqrt{c_{i,i}} \cdot \left(\nv_{i,i}\right)^{t + \frac{e}{\epsilon_i \epsilon_j (e-1)}}  \notag \\
  & = \sqrt{c_{i,i}} \cdot \left(\nv_{i,i}\right)^{t + 2 \gamma},
	\end{align} 
	where $\gamma = \frac{e}{2\epsilon_i \epsilon_j (e-1)}$  was also  defined in Lemma~\ref{prop:average_two_rounds_lb}. 
	
	\medskip 
	
	By definition of the normalized matrix $\nv$ in  (\ref{eq:rescaling_damped}), we have  $\nv_{i,j} = {v_{i,j}}/{\delta}$ $\forall i,j \in [n]$, where $\delta = \max_{k,\ell} v_{k,\ell}$. 
	Recall 
	$\nx_i(t) = \frac{\dmx_i(t)}{\sigma \delta^{t}}$ and $\ny_{i,j}(t) = \dmy_{i,j}(t)$ for each $i,j \in [n]$, where   $\sigma = \sum_{k} x_{k}(0)$ (by Lemma~\ref{lem:normalized_system_relation}). 
	Substituting these identities in  (\ref{eq:lb_nxi_alone}) gives 
	\begin{align}  \label{eq:inequality_amounts_self_loops}
		\left(\frac{\dmx_i(t)}{\sigma \delta^t} \right) \cdot \dmy_{i,i}(t)^{\frac{\epsilon_i + e^{-t}}{\epsilon_i \epsilon_i + e^{-t}}}  & \geq \sqrt{c_{i,i}} \cdot \left(\frac{v_{i,i}}{\delta}\right)^{t + 2 \gamma} \iff \notag \\
		\dmx_i(t) & \geq \sigma \sqrt{c_{i,i}} \cdot \left(\frac{ v_{i,i}}{\delta}\right)^{2 \gamma} \cdot \left(v_{i,i}\right)^{t} \,.
	\end{align} 
	Let $\widetilde{a}_i = \sqrt{c_{i,i}} \cdot \left(\frac{ v_{i,i}}{\delta}\right)^{2 \gamma}$. Then inequality (\ref{eq:inequality_amounts_self_loops}) is equivalent to
	 $\dmx_i(t) \geq \widetilde{a}_i \cdot \left(v_{i,i}\right)^{t}$ ${\textbf{({i})}}$.
	
	Lemma~\ref{prop:average_two_rounds_lb} ensures the existence of constants  $\widetilde{b}_{i,j} > 0$ $\forall j \in [n]$, so that $$
	\frac{1}{2}\left({\dmx_i(t)} + {\dmx_i(t+1)}\right) 
	\geq   \widetilde{b}_{i,j} \cdot \left({v_{i,j}} {v_{j,i}}\right)^{\frac{t}{2}}
	\;\;\; {\textbf{({ii})}}$$
	
	Then inequalities {\textbf{({i})}} and {\textbf{({ii})}} imply the theorem statement with $$\widetilde{c}_i = \min\{\widetilde{a}_i, \min_{j \in [n]} \widetilde{b}_{i,j}\}\,.$$
 This completes the proof.
\end{proof}

\end{document}